\documentclass[12pt]{article}

\usepackage{epsfig}
\usepackage{amsmath}
\usepackage{amssymb}
\usepackage{graphicx}
\usepackage{latexsym}
\usepackage{amsfonts}
\usepackage{bm}
\usepackage{textcomp}
\newcommand{\nn}{\nonumber}
\newcommand{\beq}{\begin{equation}}
\newcommand{\eeq}{\end{equation}}
\def\bea{\begin{eqnarray}}
\def\eea{\end{eqnarray}}
\def\bcen{\begin{center}}
\def\ecen{\end{center}}

\begin{document}

\renewcommand{\thefootnote}{\alph{footnote}}

\begin{flushright}
WITS-CTP-72
\end{flushright}
\vskip 0.2cm

\bcen
%%%%%%%%%%%%%%%%%%%%%%%%%%%%%%%%%%%%%
%  Title
{\bf\Large Analysis of the lepton polarisation asymmetries of ${\bar B} \to {\bar K}_2(1430) \, \ell^+ \, \ell^-$ decay}
\vskip 1cm
%%%%%%%%%%%%%%%%%%%%%%%%%%%%%%%%%%%%%
%  Authors
{\bf
S.~R.~Choudhury$^1$\footnote{src@iiserbhopal.ac.in},
A.~S.~Cornell$^2$\footnote{alan.cornell@wits.ac.za}
and J.~D.~Roussos$^2$\footnote{joe.roussos@gmail.com}
}
%%%%%%%%%%%%%%%%%%%%%%%%%%%%%%%%%%%%%
%  Affiliations
\vskip .7cm
{\sl
$^1$Indian Institute of Science Education and Research, Bhopal, India,\\
$^2$National Institute for Theoretical Physics; School of Physics, University of the Witwatersrand, Wits 2050, South Africa}
\ecen
\vskip 0.5cm
%%%%%%%%%%%%%%%%%%%%%%%%%%%%%%%%%%%%%
%  Abstract
\begin{abstract}

\noindent In this work we will study the longitudinal polarisations of both leptons in the decay process ${\bar B} \to {\bar K}_2(1430)\,  \ell^+ \, \ell^-$. This process has all the features of the related and well investigated process ${\bar B} \to {\bar K}^*(890) \, \ell^+ \, \ell^-$, with theoretically comparable branching ratios. The polarised differential decay rates as well as the single and double polarisation asymmetries are worked out, where the sensitivity of these to possible right-handed couplings for the related $b \to s$ radiative decay (and other generic BSM parameters) are also investigated.
\end{abstract}

\renewcommand{\thefootnote}{\arabic{footnote}}
\setcounter{footnote}{0}  % Resetting the footnote counter

%%%%%%%%%%%%%%%%%%%%%%%%%%%%%%%%%%%%%
%  Section 1: Introduction

\section{Introduction}\label{sec1}

\par Flavour Changing Neutral Currents (FCNCs) in weak decays provide fertile ground for testing the structure of weak interactions as these decays are forbidden at the tree level. As such, they proceed as higher order loop effects. Consequently they are sensitive to finer details of the basic interactions responsible for the process and therefore provide a natural testing ground for any theories beyond the Standard Model (SM). Of the FCNC decays the radiative mode $B \to K^*(890) \, \gamma$ has been experimentally measured, with a lot of theoretical work also having gone into its study. A related decay, $B \to K_2(1430) \, \gamma$ \cite{Aubert:2003zs} has also been observed experimentally, with branching ratios comparable to the decay $B \to K^*(890) \, \gamma$. The related decay processes with a lepton pair instead of the photon, which have already been seen for the $K^*(890)$ case, can be expected to be seen for the $K_2(1430)$ case, since the branching ratios are comparable. Analysis of this latter process will therefore be a useful complement to the much investigated analysis for the  $K^*(890) \, \ell^+ \, \ell^-$  process for confrontation with theory, since the analysis probes the effective Hamiltonian in a similar but not identical way. Data on $K_2(1430) \, \ell^+ \, \ell^-$ would thus provide an independent test of the predictions of the SM.

\par In this paper we study the angular distribution of the rare $B$-decay ${\bar B} \to {\bar K}_2(1430) \, \ell^+ \, \ell^-$ using the standard effective Hamiltonian approach and form factors that have already been estimated for the corresponding radiative decay ${\bar B} \to {\bar K}_2(1430) \, \gamma$ \cite{Cheng:2004yj}. The additional form factors for the dileptonic channel are estimated using the Large Energy Effective Theory (LEET) \cite{Charles:1998dr}, which enables one to relate the additional form factors to the form factors of the radiative mode. We note here that the LEET does not take account of collinear gluons and this deficiency is remedied in the Soft Collinear Effective theory introduced by Bauer, Fleming, Pirjol and Stewart \cite{Bauer}. However, as they have shown, interactions with collinear gluons preserve the LEET relations between the form factors for a heavy to light decay as long as we ignore terms suppressed by $m/E$, where $m$ is the mass of the light meson and $E$ is its energy in the $B$-meson's rest frame. 

\par The importance of polarization effects for the process $B \to K^* \ell^+ \ell^-$ was first pointed out by Hewett \cite{Hewett:1995dk} and subsequently by others \cite{Kruger:1996cv}. These papers considered other observables beyond the differential decay rate. We have earlier considered decay rates \cite{Choudhury:2009fz} for the process under consideration, and in this paper we shall study longitudinal polarization observables in the process. In earlier works $\tau$ lepton polarisation asymmetries were analysed in various Beyond the SM (BSM) scenarios for both inclusive and exclusive $B$ decays. Though there is no experimental data on these observables yet, these asymmetries have been observed to be extremely sensitive to the structure of new interactions, making them ideal for testing BSM physics. As such, the paper is organized as follows: In section 2 we will give the relevant effective Hamiltonian and the LEET form factors for the process under consideration. In section 3 we will work out  the expressions for the polarised differential decay rate for the semi-leptonic decay mode and the various lepton polarisation asymmetries. We will conclude with our results in section 4.

%%%%%%%%%%%%%%%%%%%%%%%%%%%%%%%%%%%%%
%  Section 2: Effective Hamiltonian and Form Factors

\section{Effective Hamiltonian and Form Factors}\label{sec2}

\par The process in which we are interested (${\bar B} \to {\bar K}_2(1430) \, \ell^+ \, \ell^-$) is governed by the quark level decay $b \to s \, \ell^+ \, \ell^-$. By integrating out the heavy degrees of freedom from the theory we obtain the effective Hamiltonian \cite{Ali:2004dq}:
\bea
{\cal H}^{eff} &=& \frac{\alpha G_F}{\sqrt{2}\pi} V_{tb} V_{ts}^* \Bigg[ -  i C_{7L} m_b \frac{q^\nu}{q^2} \left(T_{\mu \nu} + T^5_{\mu \nu}\right) L^\mu -  i C_{7R} m_b \frac{q^\nu}{q^2} \left(T_{\mu \nu} - T^5_{\mu \nu}\right) L^\mu \nonumber \\
&& \hspace{2.2cm} + {1\over 2}\left( C_9^{eff} - C_{10} \right) (V - A)_\mu
   \left(L^\mu - L_5^\mu \right) \nonumber \\
&& \hspace{2.3cm} + {1\over2}\left( C_9^{eff} + C_{10} \right) (V - A)_\mu
   \left( L^\mu - L_5^\mu \right) \Bigg] \; . \label{eqn:1}
\eea
where $L^\mu = \bar{\ell}\gamma^\mu \ell$ and $L_5^\mu = \bar{\ell} \gamma^\mu \gamma_5 \ell$ are the lepton bilinears, whilst the $C$'s are the Wilson coefficients. $C_{7R}$ in the SM is zero but may arise in models BSM. As such we will retain this in order to see its effect on some of the experimentally observable quantities. $C_9^{eff}$ includes the short-distance Wilson coefficient as well as long distance effects simulated through the lepton pair being produced by decay of $c {\bar c}$ resonances, where these are fully spelled out in appendix \ref{appendix:b}. Note also that in equation (\ref{eqn:1}) we have used the $(V-A)$ structure for the hadronic part (except for $C_7$) 
\bea
V_\mu &=&  \left(\bar{s} \gamma_\mu b\right) \,\,\, , \label{eq:4} \\
A_\mu &=&  \left(\bar{s} \gamma_\mu \gamma_5 b\right) \,\,\, , \label{eq:5} \\
T_{\mu\nu} &=&  \left( \bar{s} \sigma_{\mu \nu} b \right) \,\,\, , \label{eq:6}  \\
T^5_{\mu\nu} &=&  \left( \bar{s} \sigma_{\mu \nu} \gamma_
  5b \right) \,\,\, . \label{eqn:2}
\eea
Note that this structure doesn't change under the transformation $V \leftrightarrow - A$ and $T_{\mu \nu} \leftrightarrow T^5_{\mu \nu}$. Furthermore, we can relate the hadronic factors of $T_{\mu\nu}$ and $T^5_{\mu \nu}$ by using the identity\footnote{Where we have used the convention that $\gamma_5= i \gamma^0 \gamma^1 \gamma^2 \gamma^3$ and that $\varepsilon_{0123}=1$.}:
$$
\sigma_{\mu \nu} = - \frac{i}{2} \varepsilon^{\mu \nu \rho \delta}
\sigma_{\rho \delta} \gamma_5 \,\,\, .
$$

\par In order to enable us to study the sensitivity of our results to BSM physics, we have included a possible $C_{7R}$ in the effective Hamiltonian, which is otherwise absent in the SM. This is similar to the work of  Kim {\it et al.}\cite{Kim:2000dq} for the decay channel $B \to K^*(890) \, \ell^+ \, \ell^-$. Physics BSM often results in non-standard $Z^{\prime}$ coupling to quarks. As far as the effective Hamiltonian is concerned, this results in modifying the values of  $C_9$ and $C_{10}$ away from their SM values. Following, a recent study of such deviations and the constraints imposed on them from known experimental data \cite{Chiang}, we write additive complements to both $C_9$ and $C_{10}$ that we will detail later.

\par We now define the hadronic form factors for the ${\bar B} \to {\bar K}_2(1430)$ decay as:
\bea
\langle K_2 (p') | V_\mu  | B (p) \rangle &=& 2  V  \epsilon^{* \alpha \beta} \varepsilon_{\alpha \mu \nu \rho} pÍ^{\nu} p^{\rho} p_{\beta} \,\,\, ,  \label{eqn:3} \\
\langle K_2 (p') | A_\mu | B (p) \rangle &=& \epsilon^{* \alpha \beta}  \Big[ 2 A_1 g_{ \alpha\mu} p_{\beta}                                                         +  A_2  p_{\alpha} p_{\beta}p_{\mu}  +  A_3  p_{\alpha} p_{\beta} p'_{\mu} \Big] \,\,\, , \label{eqn:4}\\
\langle K_2 (p') |  i   T_{\mu \nu}  q^{\nu} | B (p) \rangle &=&  \frac{2 i U_1}{m_B}  \epsilon^{* \alpha \beta} \varepsilon_{\mu\alpha\lambda\rho} p_{\beta}  p^{\lambda}  p'^{\rho} \,\,\, , \label{eqn:5}   \\
\langle K_2 (p') |  i  T^5_{\mu\nu} q^{\nu} | B (p) \rangle &=&  \epsilon^{* \alpha \beta}   \left( \frac{U_2 (p+p')_{\beta} }{m_B} \right)  \Bigg [  g_{\mu\alpha} (p +p'). q  - (p+p')_{\mu} q^{\alpha} \Bigg]  \nonumber \\
&&-  \epsilon^{* \alpha \beta} p_\alpha p_\beta \Bigg[ q_{\mu} - (p+p')_{\mu} \frac{q^2}{(p+p').q}\Bigg] \frac{U_3}{m_B} \,\,\, , \label{eqn:6}
\eea
where $\epsilon^{* \alpha \beta}$ is the polarisation vector for the $K_2$.

\par This leads to a matrix element:
\bea
{\cal M} & = & \left( \frac{\alpha G_F \lambda_{CKM}}{2 \sqrt{2} \pi} \right) \epsilon^{* \alpha \beta} \Bigg[ (L^\mu) H^V_{\mu \alpha \beta} + ( L_5^\mu ) H^A_{\mu \alpha \beta} \Bigg] \,\,\, ,
\eea
where
\bea
H^A_{\mu \alpha \beta} & = & C_{10} \Bigg[ 2 V \varepsilon_{\alpha \mu \nu \rho} pÍ^{\nu} p^{\rho} p_{\beta} - 2 A_1 g_{ \alpha\mu} p_{\beta} -  A_2  p_{\alpha} p_{\beta}p_{\mu}  -  A_3  p_{\alpha} p_{\beta} p'_{\mu} \Bigg] \,\,\, , \label{eqn:7} \\
H^V_{\mu \alpha \beta} & = & C_9^{eff} \Bigg[ 2 V \varepsilon_{\alpha \mu \nu \rho} pÍ^{\nu} p^{\rho} p_{\beta} - 2 A_1 g_{ \alpha\mu} p_{\beta} -  A_2  p_{\alpha} p_{\beta}p_{\mu}  -  A_3  p_{\alpha} p_{\beta} p'_{\mu} \Bigg] \nn \\
&& - 2 (C_{7L} + C_{7R}) \frac{m_b}{q^2} \times \left( \frac{2 i U_1}{m_B} \varepsilon_{\mu\alpha\lambda\rho} p_{\beta}  p^{\lambda}  p'^{\rho} \right) \nn \\
&& + 2 (C_{7L} - C_{7R}) \frac{m_b}{q^2} \times \left( \frac{U_2 (p+p')_{\beta} }{m_B} \right)  \Bigg [  g_{\mu\alpha} (p +p'). q  - (p+p')_{\mu} q^{\alpha} \Bigg] \nn \\
&& + 2 (C_{7L} - C_{7R}) \frac{m_b}{q^2} \times \left( p_\alpha p_\beta \Bigg[ q_{\mu} - (p+p')_{\mu} \frac{q^2}{(p+p').q}\Bigg] \frac{U_3}{m_B} \right) \; . \label{eqn:8}
\eea

\par The helicity states for the $K_2$ are:
\begin{eqnarray}
\epsilon^{* \alpha \beta}(+2) & = & \epsilon^\alpha_{(+)} \epsilon^\beta_{(+)} \nonumber \\
\epsilon^{* \alpha \beta}(+1) & = & \frac{1}{\sqrt{2}} \left(\epsilon^\alpha_{(+)} \epsilon^\beta_{(0)} + \epsilon^\alpha_{(0)} \epsilon^\beta_{(+)} \right)  \nonumber \\
\epsilon^{* \alpha \beta}(+2) & = & \frac{1}{\sqrt{6}} \left(\epsilon^\alpha_{(+)} \epsilon^\beta_{(-)} + \epsilon^\alpha_{(-)} \epsilon^\beta_{(+)} \right) + \sqrt{\frac{2}{3}} \epsilon^\alpha_{(0)} \epsilon^\beta_{(0)} \nonumber \\
\epsilon^{* \alpha \beta}(-1) & = & \frac{1}{\sqrt{2}} \left(\epsilon^\alpha_{(-)} \epsilon^\beta_{(0)} + \epsilon^\alpha_{(0)} \epsilon^\beta_{(-)} \right)  \nonumber \\
\epsilon^{* \alpha \beta}(-2) & = & \epsilon^\alpha_{(-)} \epsilon^\beta_{(-)} \; ,  \label{eqn:9} 
\end{eqnarray}
where
\begin{eqnarray}
\big[\epsilon_{(+)}^\mu \big] & = & \frac{1}{\sqrt{2}}(0, -i, \cos\theta, - \sin\theta ) \nonumber \\
\big[\epsilon^\mu_{(0)} \big] & = & \frac{1}{m_K}(k, 0, E_K\sin\theta, E_K \cos\theta ) \nonumber \\
\big[ \epsilon_{(-)}^\mu \big] & = & \frac{1}{\sqrt{2}}(0, -i, -\cos\theta, \sin\theta )\; ,  \label{eqn:10} 
\end{eqnarray}
and the lepton bilinears $L^\mu$ and $L_5^\mu$ are given by (for lepton helicity $(\lambda_1, \lambda_2)$):
\begin{eqnarray}
\big[L^\mu (+,+)\big] & = & (0, 0, 0, 1) \nonumber \\
\big[L^\mu (+,-)\big] & = & (0, E_\ell/m_\ell, i E_\ell/m_\ell, 0) \nonumber \\
\big[L^\mu (-,+) \big] & = & (0, E_\ell/m_\ell, -i E_\ell/m_\ell, 0) \nonumber \\
\big[L^\mu (-,-) \big] & = & (0, 0, 0, -1) \nonumber \\
\big[L_5^\mu (+,+) \big] & = & (1, 0, 0, 0) \nonumber \\
\big[L_5^\mu (+,-) \big] & = & (0, -p_\ell/m_\ell, -i p_\ell/m_\ell, 0) \nonumber \\
\big[L_5^\mu (-,+) \big] & = & (0, p_\ell/m_\ell, -i p_\ell/m_\ell, 0) \nonumber \\
\big[L_5^\mu (-,-) \big] & = & (1, 0, 0, 0) \; . \label{eqn:11}
\end{eqnarray}

\par The form factors introduced in equations (\ref{eqn:3}--\ref{eqn:6}) can be related using the LEET approach, that is, using equations (44--48) of J. Charles {\it et al.} \cite{Charles:1998dr} we get:
\bea
V &=& \frac {i  A_1} { m_B  E} \; , \nonumber \\
V&=& - \frac{i U_1}{m_B^2} \; , \nonumber \\
U_2 &=& - 2 A_1 \; , \nonumber \\
A_2 &=&0\;  , \nonumber \\
A_3&=& \frac{2  U_3}{m_B^2}\;  , \label{eqn:12}
\eea
where we have taken the limit of the heavy quark mass going to infinity and $E = p . p'/m_B$. Note that with this approach we have introduced no extra hadronic form factors beyond what is required for the radiative mode. Thus, once we are able to describe the radiative mode we have in effect a check on the model from the dileptonic mode. The radiative mode form factors $U_1$, $U_2$ and $U_3$ have been  given by Cheng and Chua \cite{Cheng:2004yj} in their analysis of radiative charmless decays of the $B$-meson using covariant light cone wave functions. We shall use their results.

%%%%%%%%%%%%%%%%%%%%%%%%%%%%%%%%%%%%%
%  Section 3: Kinematics and the lepton polarisation asymmetries

\section{Kinematics and the lepton polarisation asymmetries}\label{sec:3}

\par If we now use the dilepton centre of mass (CM) frame, where $\theta$ shall be the angle between the $K_2$ meson and the $\ell^+$, and $s$ is the energy squared of the outgoing leptons, then we can write our polarised differential decay widths as:
\begin{eqnarray}
\frac{d \Gamma_{ij}^k}{ds d(\cos\theta)} & = & (2 m_\ell)^2 \frac{\alpha^2 G_F^2}{2^{12} \pi^5 m_B^3} \left| V_{tb} V_{ts}^* \right|^2 \sqrt{1 - \frac{4 m_\ell^2}{s}} \lambda^{1/2} \left| {\cal M}_{ij}^k \right|^2 \; , \label{eqn:13}
\end{eqnarray}
where we have writen our amplitudes as ${\cal M}_{ij}^k$ with $i$ as the $\ell^+$ helicity, $j$ as the $\ell^-$ helicity and $k$ the $K_2$'s helicity, where all $k=\pm2$ were found to be zero:
\begin{eqnarray}
{\cal M}_{++}^\pm & = & \sin\theta \left( C_{7L} (\mp H_2 - H_1) + C_{7R} (\mp H_2 - H_1) \mp C_9 (H_3 \pm H_4) \right) \; ,\nonumber \\
{\cal M}_{+-}^\pm & = & \frac{1}{2} \left( 1 \mp \cos\theta \right) \left( \frac{i \sqrt{s}}{m_\ell}\right) \Bigg(  C_{7L} ( \mp H_1 +  H_2) +  C_{7R}  (- H_2 \mp  H_1) \nonumber \\
&& \hspace{4cm} + \sqrt{1 - \frac{4 m_\ell^2}{s}} C_{10}  ( H_3 \pm  H_4) -  C_9 (H_3 \pm H_4) \Bigg) \; , \nonumber \\ 
{\cal M}_{-+}^\pm & = & \frac{1}{2} \left( 1 \pm \cos\theta \right) \left( \frac{i \sqrt{s}}{m_\ell}\right) \Bigg(  C_{7L} ( - H_1 + H_2) +  C_{7R}  (-H_2 -  H_1) \nonumber \\
&& \hspace{4cm}  + \sqrt{1 - \frac{4 m_\ell^2}{s}} C_{10}  ( - H_3 \mp  H_4)  - C_9 (H_3 + H_4) \Bigg) \; , \nonumber \\ 
{\cal M}_{--}^\pm & = & \sin\theta \left( C_{7L} (\mp H_2 - H_1) + C_{7R} (\pm H_2 - H_1) + C_9 (\pm H_3 - H_4) \right) \; , \nonumber \\
{\cal M}_{\pm \pm}^0 & = & \cos\theta \left( \pm(C_{7L}-C_{7R})(Z_1 + Z_2) \pm C_9 Y_2 \right) + Y_1 C_{10} \; , \nonumber \\
{\cal M}_{\pm \mp}^0 & = & \sin\theta \left( \frac{i \sqrt{s}}{m_\ell}\right) \left(\pm Y_2 C_9 \pm (C_{7L}-C_{7R})(Z_1+Z_2) - \sqrt{1 - \frac{4 m_\ell^2}{s}} C_{10} Y_2 \right) \; , \label{eqn:14}
\end{eqnarray}
where
\begin{eqnarray}
H_1 & = & \frac{m_b \lambda U_1}{2 m_B m_K s} \nonumber \\
H_2 & = & \frac{m_b \lambda^{1/2} \left( m_B^2 - m_K^2 \right) U_2}{2 m_B m_K s} \nonumber \\
H_3 & = & \frac{2 \lambda^{1/2} A_1}{4 m_K} \nonumber \\
H_4 & = & \frac{i V \lambda}{4 m_K} \nonumber \\
Z_1 & = & \frac{m_b \lambda^{1/2}}{\sqrt{6} m_K^2 m_B s^{3/2}} U_2 \left( \lambda -  (m_B^2 - m_K^2) (m_B^2 - m_K^2 - s) \right) \nonumber \\
Z_2 & = & \frac{m_b \lambda^{3/2} U_3}{\sqrt{6} (m_B^3 m_K^2 - m_B m_K^4) \sqrt{s}} \nonumber \\
Y_1 & = & \frac{\lambda}{4 \sqrt{6} m_K^2 \sqrt{s}} \left( -4 A_1 - (m_B^2 - m_K^2) (A_2 + A_3) + (A_3 - A_2)s \} \right) \nonumber \\
Y_2 & = & \frac{\lambda^{1/2}}{4 \sqrt{6} m_K^2 \sqrt{s}} \left( (A_2 + A_3)\lambda + 4 A_1 (m_B^2 - m_K^2-s) \right) \nonumber \\
\mathrm{and} \; \lambda & = & m_B^4 + m_K^4 + s^2 - 2 m_B^2 m_K^2 - 2 s m_B^2 - 2 s m_K^2 \; . \label{eqn:15}
\end{eqnarray}

\par Equipped with the above we can now define the various single lepton and double lepton polarisation asymmetries. The single lepton longitudinal polarisation asymmetries are defined as:
\bea
{\cal P}_{\ell^+}^\pm & = & \Bigg[ \pm \left( \frac{d\Gamma_{++}^+}{ds}+\frac{d\Gamma_{-+}^+}{ds} + \frac{d\Gamma_{++}^0}{ds}+\frac{d\Gamma_{-+}^0}{ds} + \frac{d\Gamma_{++}^-}{ds}+\frac{d\Gamma_{-+}^-}{ds} \right) \nonumber \\
&& \hspace{0.5cm} \mp \left( \frac{d\Gamma_{+-}^+}{ds}+\frac{d\Gamma_{--}^+}{ds} + \frac{d\Gamma_{+-}^0}{ds}+\frac{d\Gamma_{--}^0}{ds} + \frac{d\Gamma_{+-}^-}{ds}+\frac{d\Gamma_{--}^-}{ds} \right)\Bigg] \Bigg/ \frac{d\Gamma_{tot}}{ds}  \; , \nonumber \\
{\cal P}_{\ell^-}^\pm & = & \Bigg[ \pm \left( \frac{d\Gamma_{++}^+}{ds}+\frac{d\Gamma_{+-}^+}{ds} + \frac{d\Gamma_{++}^0}{ds}+\frac{d\Gamma_{+-}^0}{ds} + \frac{d\Gamma_{++}^-}{ds}+\frac{d\Gamma_{+-}^-}{ds} \right) \nonumber \\
&& \hspace{0.5cm} \mp \left( \frac{d\Gamma_{-+}^+}{ds}+\frac{d\Gamma_{--}^+}{ds} + \frac{d\Gamma_{-+}^0}{ds}+\frac{d\Gamma_{--}^0}{ds} + \frac{d\Gamma_{-+}^-}{ds}+\frac{d\Gamma_{--}^-}{ds} \right)\Bigg] \Bigg/\frac{d\Gamma_{tot}}{ds} \; . \label{eqn:16}
\eea
Along the same lines we can also define the double lepton polarisation asymmetries:
\bea
{\cal P}_{\ell^+ \ell^-}^{\pm\pm} & = & \Bigg\{ \left[ \left( \frac{d\Gamma_{++}^+}{ds} + \frac{d\Gamma_{++}^0}{ds} + \frac{d\Gamma_{++}^-}{ds} \right) \mp \left( \frac{d\Gamma_{+-}^+}{ds} + \frac{d\Gamma_{+-}^0}{ds} + \frac{d\Gamma_{+-}^-}{ds}  \right) \right] \nonumber \\
&& \hspace{0.5cm} \mp \left[ \left( \frac{d\Gamma_{-+}^+}{ds} + \frac{d\Gamma_{-+}^0}{ds} + \frac{d\Gamma_{-+}^-}{ds}  \right) \mp \left( \frac{d\Gamma_{--}^+}{ds} + \frac{d\Gamma_{--}^0}{ds} + \frac{d\Gamma_{--}^-}{ds}  \right) \right] \Bigg\} \Bigg/\frac{d\Gamma_{tot}}{ds}\; , \nonumber \\
{\cal P}_{\ell^+ \ell^-}^{\pm\mp} & = & \Bigg\{ \left[ \left( \frac{d\Gamma_{++}^+}{ds} + \frac{d\Gamma_{++}^0}{ds} + \frac{d\Gamma_{++}^-}{ds} \right) \mp \left( \frac{d\Gamma_{+-}^+}{ds} + \frac{d\Gamma_{+-}^0}{ds} + \frac{d\Gamma_{+-}^-}{ds}  \right) \right] \nonumber \\
&& \hspace{0.5cm} \pm \left[ \left( \frac{d\Gamma_{-+}^+}{ds} + \frac{d\Gamma_{-+}^0}{ds} + \frac{d\Gamma_{-+}^-}{ds}  \right) \mp \left( \frac{d\Gamma_{--}^+}{ds} + \frac{d\Gamma_{--}^0}{ds} + \frac{d\Gamma_{--}^-}{ds}  \right) \right]\Bigg\} \Bigg/\frac{d\Gamma_{tot}}{ds} \; .  \label{eqn:17}
\eea
Note that in these asymmetries we have divided by the total differential decay width 
$$\displaystyle \frac{d\Gamma_{tot}}{ds} = \sum_{ijk} \frac{d\Gamma^k_{ij}}{ds}\; . $$

%%%%%%%%%%%%%%%%%%%%%%%%%%%%%%%%%%%%%
%  Section 4: Results and Conclusion

\section{Results and Conclusion}\label{sec:4}

\par We have followed reference \cite{Kim:2000dq} for the form of the parameterisation of $C_{7L}$ and $C_{7R}$, which automatically takes care of the constraints imposed by experimental data on the radiative decay. Also, it can take into account the possibility that the phase of this term from the SM value ($u=v=0$), although present in the pure radiative decay, would not show up:
\bea
C_{7L} & = & - \sqrt{0.081} \cos x \exp\left( i (u+v)\right) \,\,\, , \nonumber \\
C_{7R} & = & - \sqrt{0.081} \sin x \exp\left( i (u-v)\right) \,\,\, . \label{eqn:18}
\eea
To also take into account possible BSM effects on the other SM Wilson coefficients, we write \cite{Chiang}:
\bea
C_9 &=& C_9^{SM} + z \,\,\, , \\
C_{10} &=& -4.546 + y  \,\,\, ,
\eea
where $C_9^{SM}$ is defined in appendix \ref{appendix:b}. $z$ and $y$ above are constrained from radiative and dileptonic decay data by \cite{Chiang}:
\bea
z < - 4.344 \,\,\, ,&& \nonumber \\
y > + 4.669 \,\,\, , && \nonumber \\
1.05 ( z + 4.01 )^2 + 1.05 (y + 4.669)^2 &<& 59.58 \,\,\, , \nonumber\\
0.61 ( z + 3.89 )^2 + 0.61 ( y + 4.669 )^2 &>& 6.38 \,\,\,  .
\eea
This way of parametrizing possible deviations from SM values takes into account the following facts: The decay rates for radiative and dileptonic $K^*(890)$ decay modes of $B$-mesons are in reasonable agreement with SM values. However, the most significant deviation from SM predictions seems to be in  the recent data of the FB asymmetry for the decay $B \to  K^*(890) \, \ell^+ \, \ell^-$ \cite{Wei:2009zv}. Within the context of the effective Hamiltonian, equation (1), such deviations can be accommodated, without significantly disturbing the predictions for the decay rates, by changing the relative signs of the Wilson co-efficients $C_9$ and $C_{10}$ relative to $C_7$. The parametrization above does just that.
	
%%%%%%%%%%%%%%%%%%%%%%%%%%%%%%%%%%%%%
\begin{table}[t]
\begin{tabular}{|c|c|c|c|c|c|}
\hline
$(x,u,v)$ & $(0,0,0)$ & $(\pi/4, 0,0)$ & $(-\pi/4, 0, 0)$ & $(\arctan (0.5), 0, 0)$ & $(\pi/2, 0, 0)$ \\
\hline
$C_{7R}/C_{7L}$ & 0 & - 1 & + 1 & 0.5 & $C_{7L}/C_{7R} = 0$ \\
\hline
$s$(for ${\cal P}_{\mu^+}^+ = 0$) (GeV$^2$) &&&&&\\
$y=z=0$ & -- & 1.7891 & 0.2925 & -- & 0.5448 \\
$y=5$, $z=-10$ & -- & -- & 0.1779 & -- & 2.2844 \\
$y=5$, $z=7.5$ & 2.1747 & 1.9146 & 2.3095 & 2.3330 & 1.5243 \\
$y=10$, $z=-5$ & 8.0627 & 8.5000 & -- & 6.7992 & -- \\
\hline
$s$(for ${\cal P}_{\mu^-}^+ = 0$) (GeV$^2$) &&&&&\\
$y=z=0$ & -- & 1.7911 & -- & -- & 0.5449 \\
$y=5$, $z=-10$ & -- & 0.1869 & -- & -- & 2.2844 \\
$y=5$, $z=7.5$ & 2.0739 & 1.8835 & 2.2496 & 2.2507 & 1.5243 \\
$y=10$, $z=-5$ & 8.0638 & 8.5003 & 1.5142 & 6.8028 & -- \\
\hline
\end{tabular}
\caption{\sl Zeroes of the muon polarisation asymmetries for $s>4 m_\mu^2$ and below $s\lesssim9.6 (GeV)^2$. Several values of the BSM parameters $x$, $y$, $z$ are presented (where we consider only $u=v=0$).}
\label{tab:one}
\end{table}
%%%%%%%%%%%%%%%%%%%%%%%%%%%%%%%%%%%%%

\par
\begin{figure}[tb]
\begin{center}
\epsfig{file=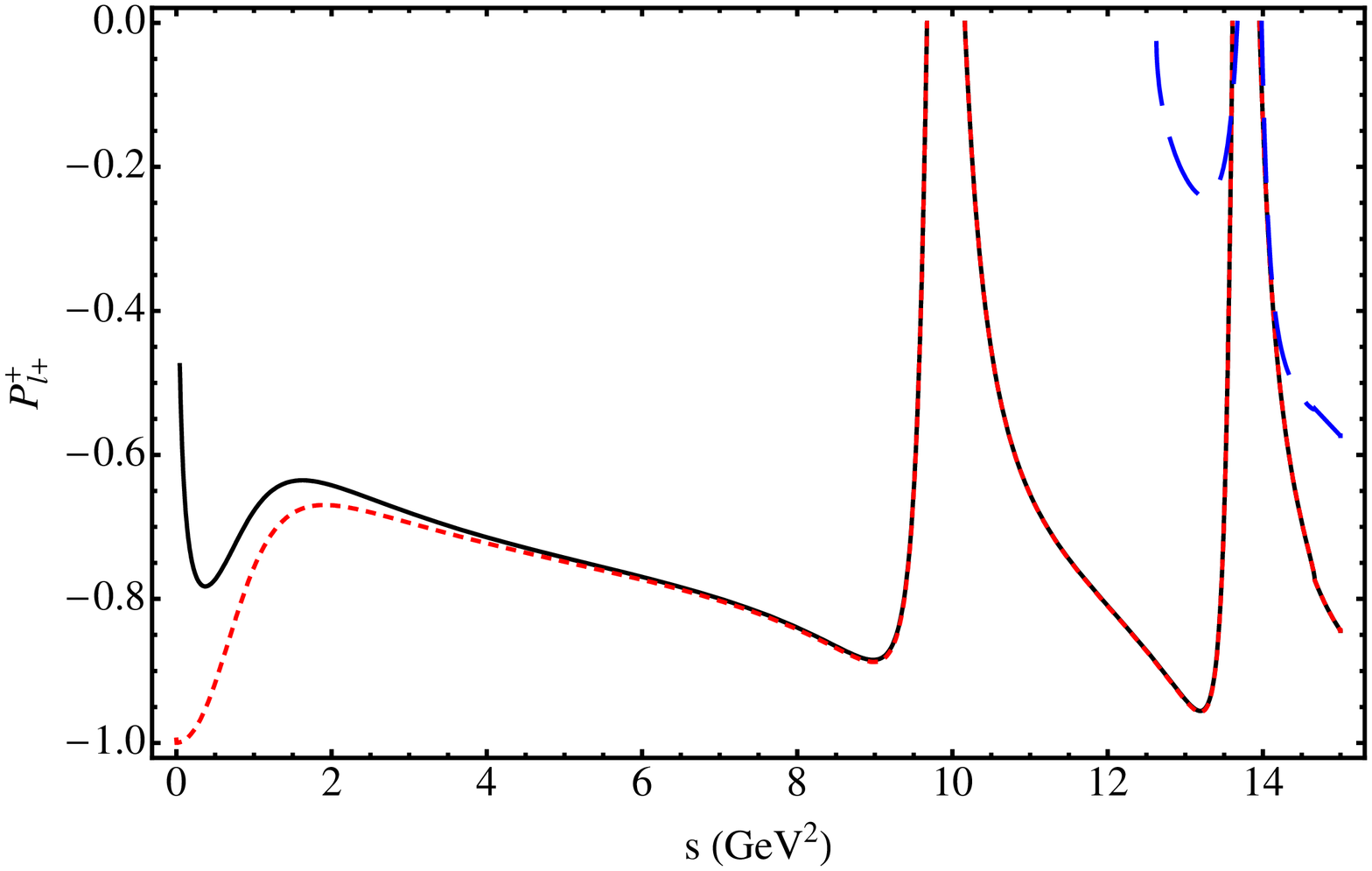,width=.45\textwidth}
\epsfig{file=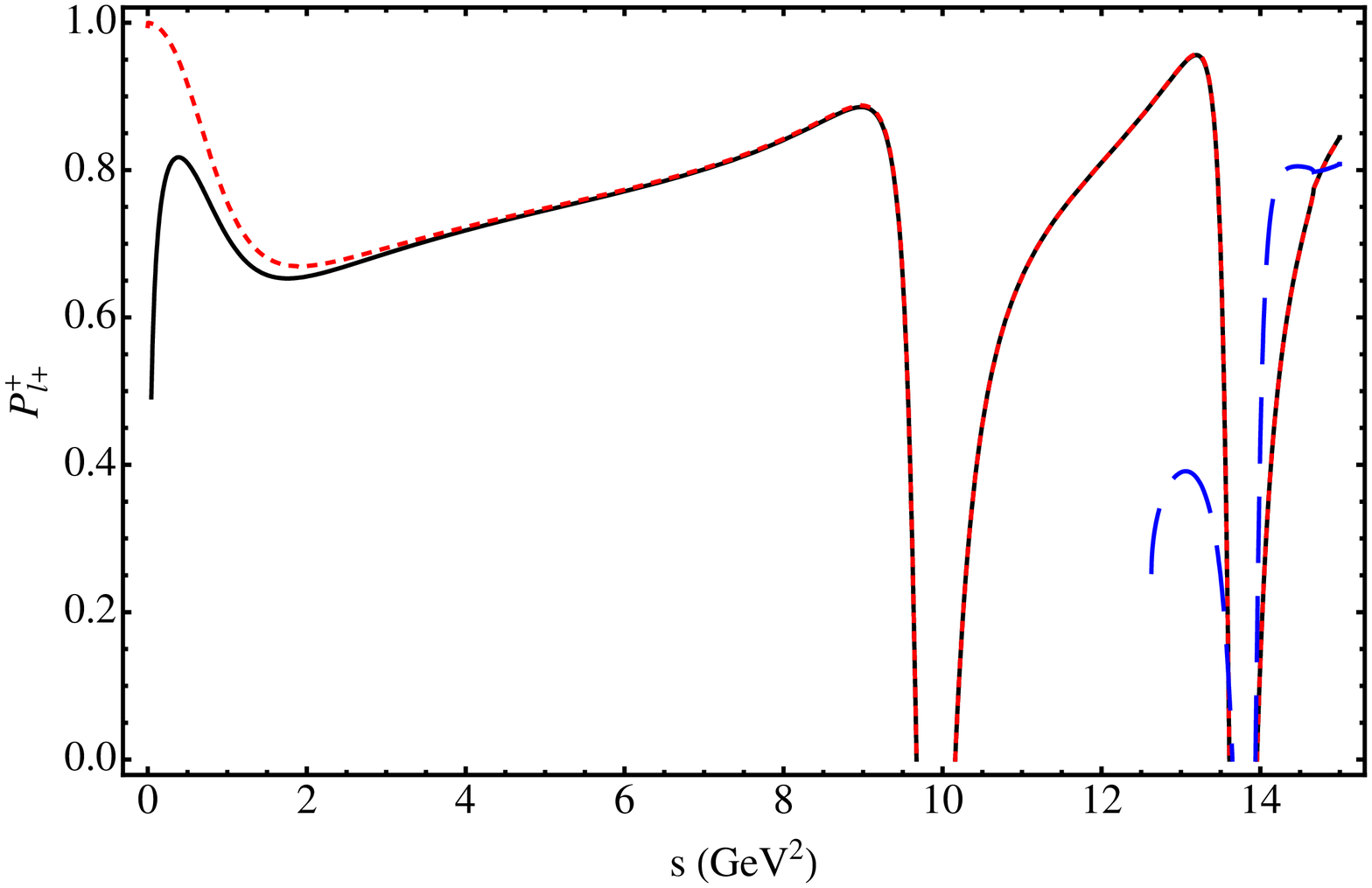,width=.45\textwidth}
\caption{\sl (Colour online) The single lepton polarisation asymmetries (${\cal P}_{\ell^+}^\lambda$) for a range of values for the three lepton species: (Black) solid line for muons, (Red) dotted line for electrons, and (Blue) dashed line for tauons. The left panel is for $\lambda = +$, the right panel $\lambda=-$, these are for $x=u=v=y=z=0$.}
\label{fig:1}
\end{center}
\end{figure}

\begin{figure}[tb]
\begin{center}
$y=0$, $z=0$ \hspace{.3\textwidth} $y=+5$, $z=-10$\\
\epsfig{file=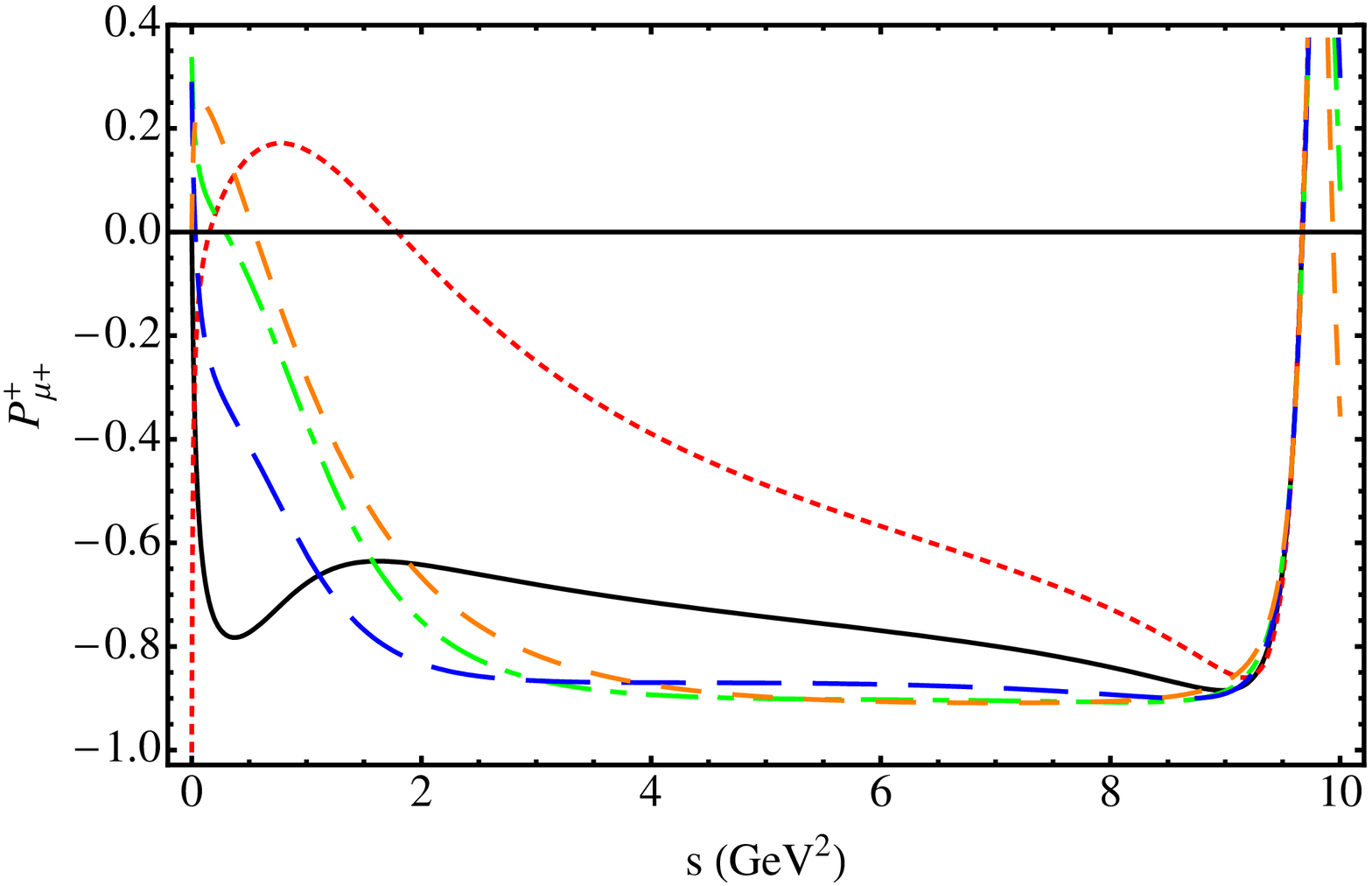,width=.45\textwidth}
\epsfig{file=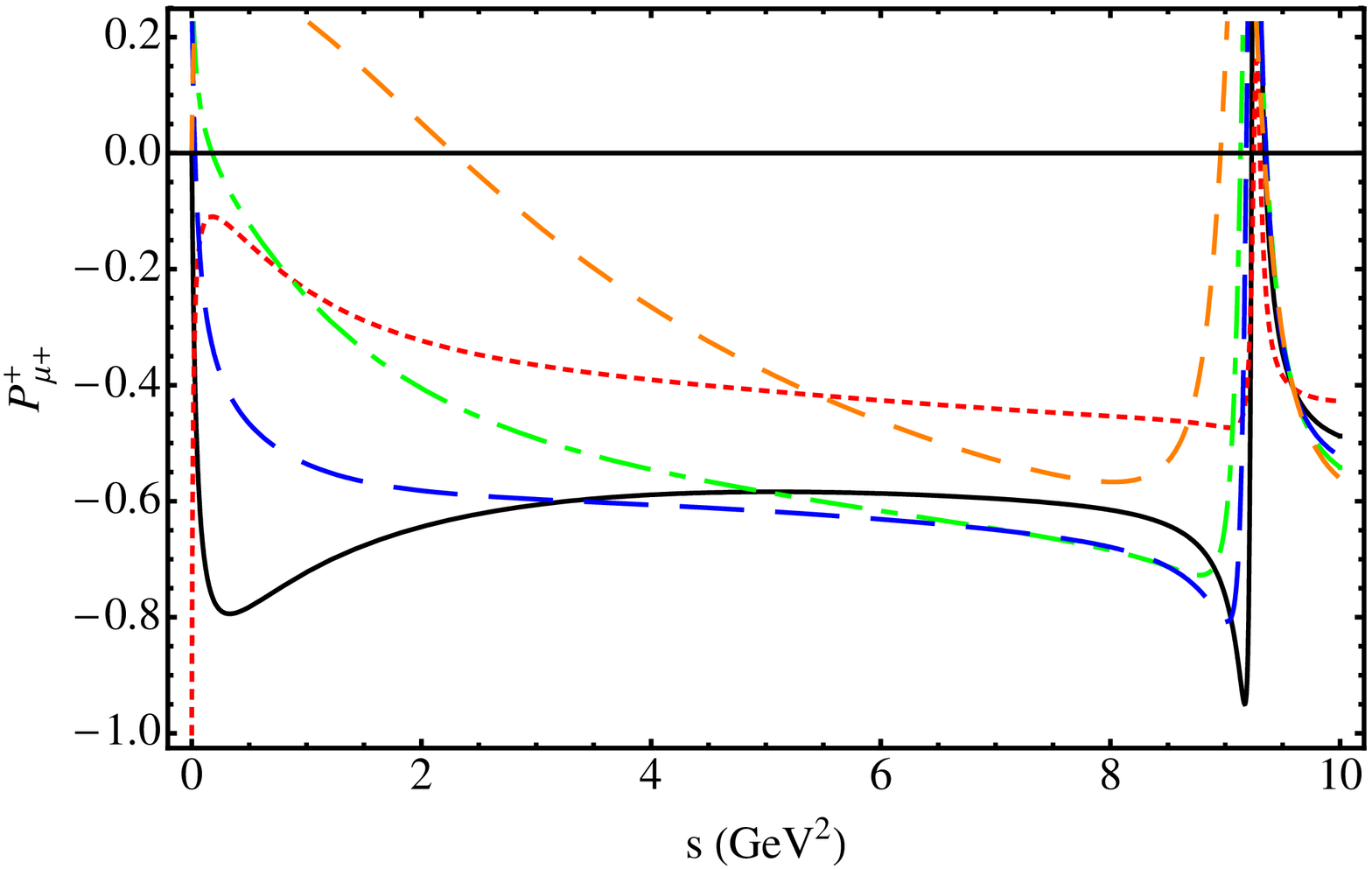,width=.45\textwidth}\\
$y=+5$, $z=+7.5$ \hspace{.3\textwidth} $y=+10$, $z=-5$\\
\epsfig{file=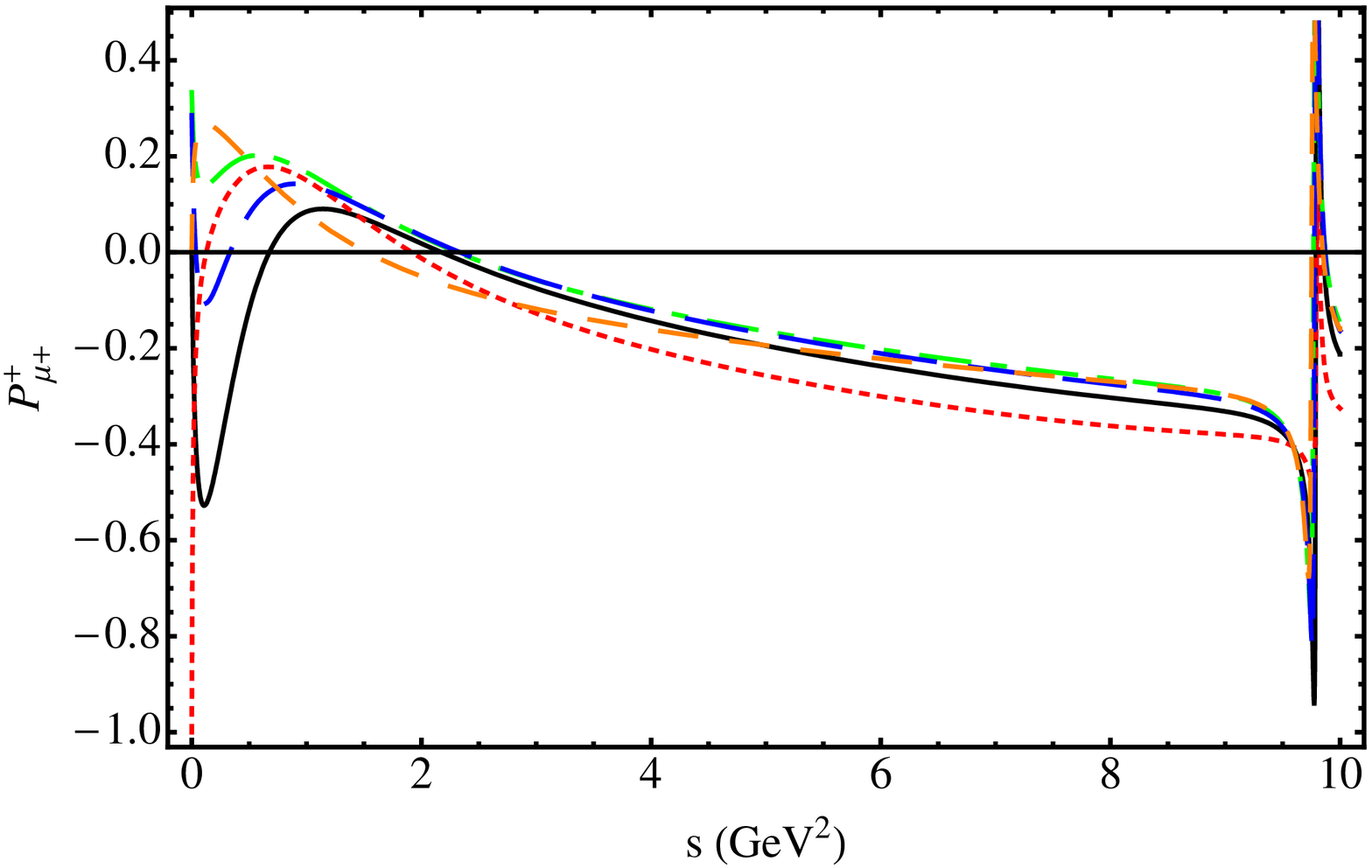,width=.45\textwidth}
\epsfig{file=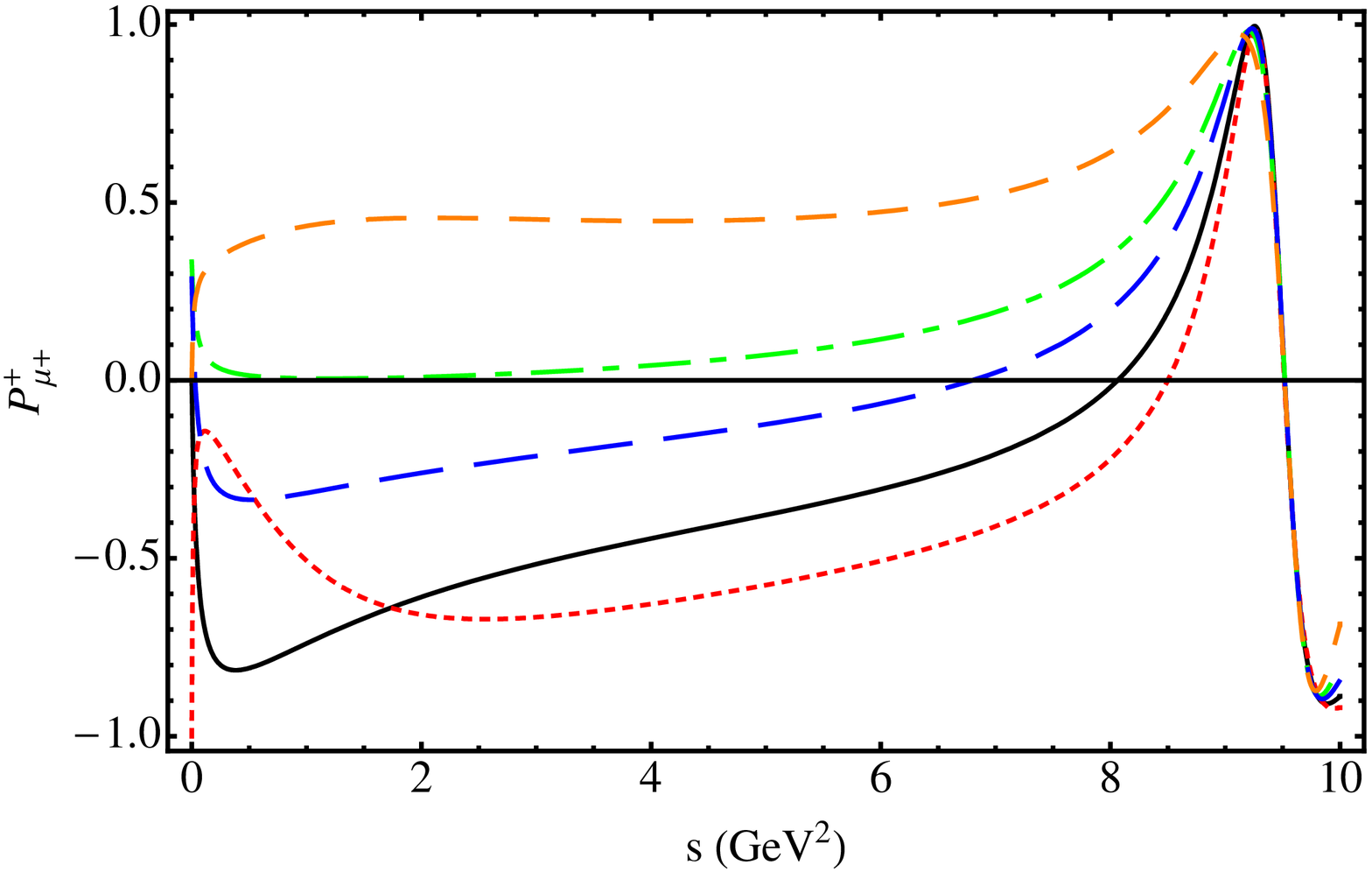,width=.45\textwidth}
\caption{\sl (Colour online) The normalised single muon polarisation asymmetry (${\cal P}_{\mu^+}^+$) for a range of values for $C_{7L}$ and $C_{7R}$.: $C_{7L}/C_{7R} = 0$ (Black) solid line, $C_{7L}/C_{7R} = +1$ (Red) dotted line, $C_{7L}/C_{7R} = -1$ (Green) dot-dashed line, $C_{7L}/C_{7R} = 0.5$ (Blue) dashed line, and $C_{7R}/C_{7L} = 0$ (Orange) small dashed line. The top left panel is for $y = 0$, $z = 0$, the top right panel for $y = +5$, $z = -10$, whilst the bottom left panel is for $y= +5$, $z = -7.5$ and the bottom right panel is for $y = +10$ and $z= -5$.}
\label{fig:2}
\end{center}
\end{figure}

\begin{figure}[tb]
\begin{center}
$y=0$, $z=0$ \hspace{.3\textwidth} $y=+5$, $z=-10$\\
\epsfig{file=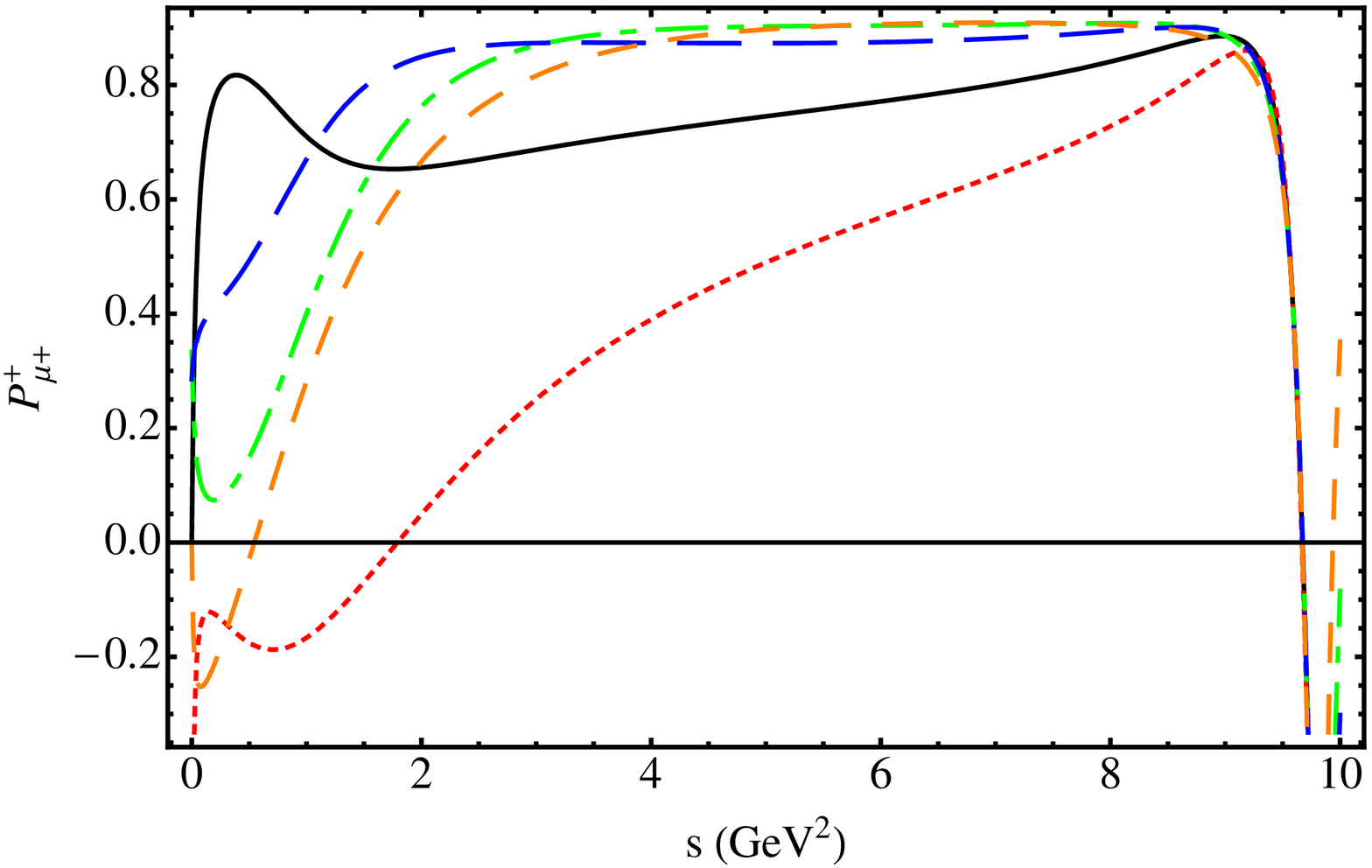,width=.45\textwidth}
\epsfig{file=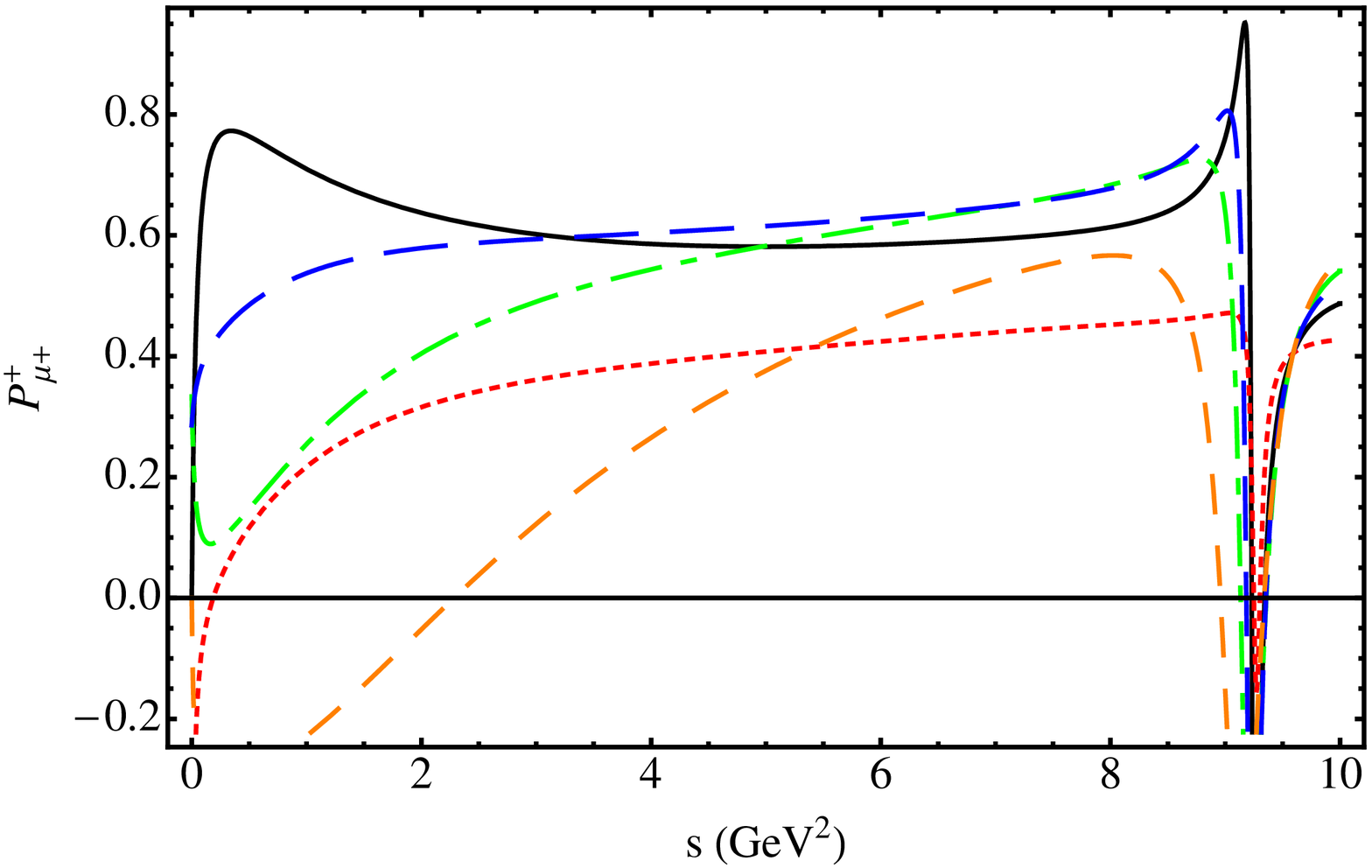,width=.45\textwidth}\\
$y=+5$, $z=+7.5$ \hspace{.3\textwidth} $y=+10$, $z=-5$\\
\epsfig{file=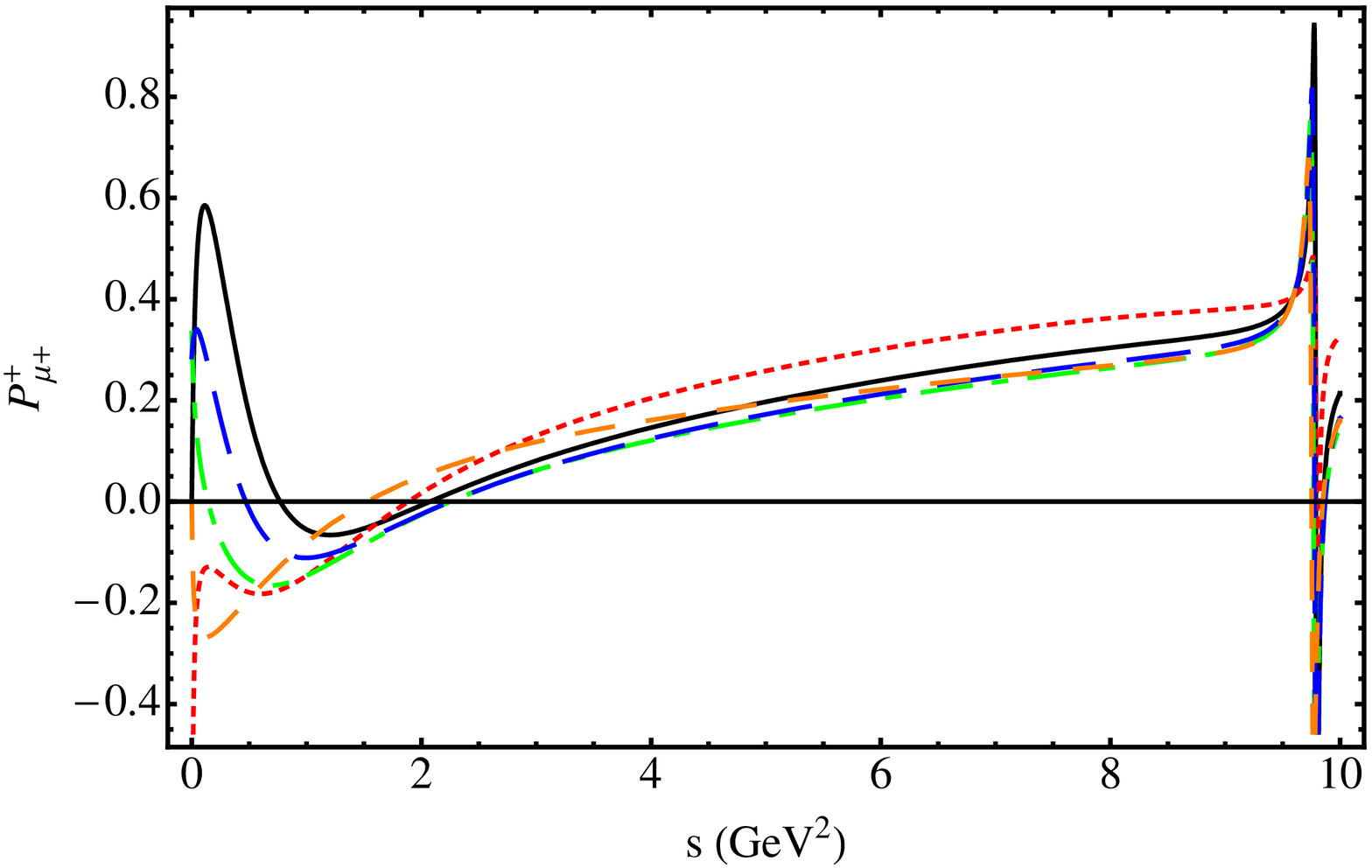,width=.45\textwidth}
\epsfig{file=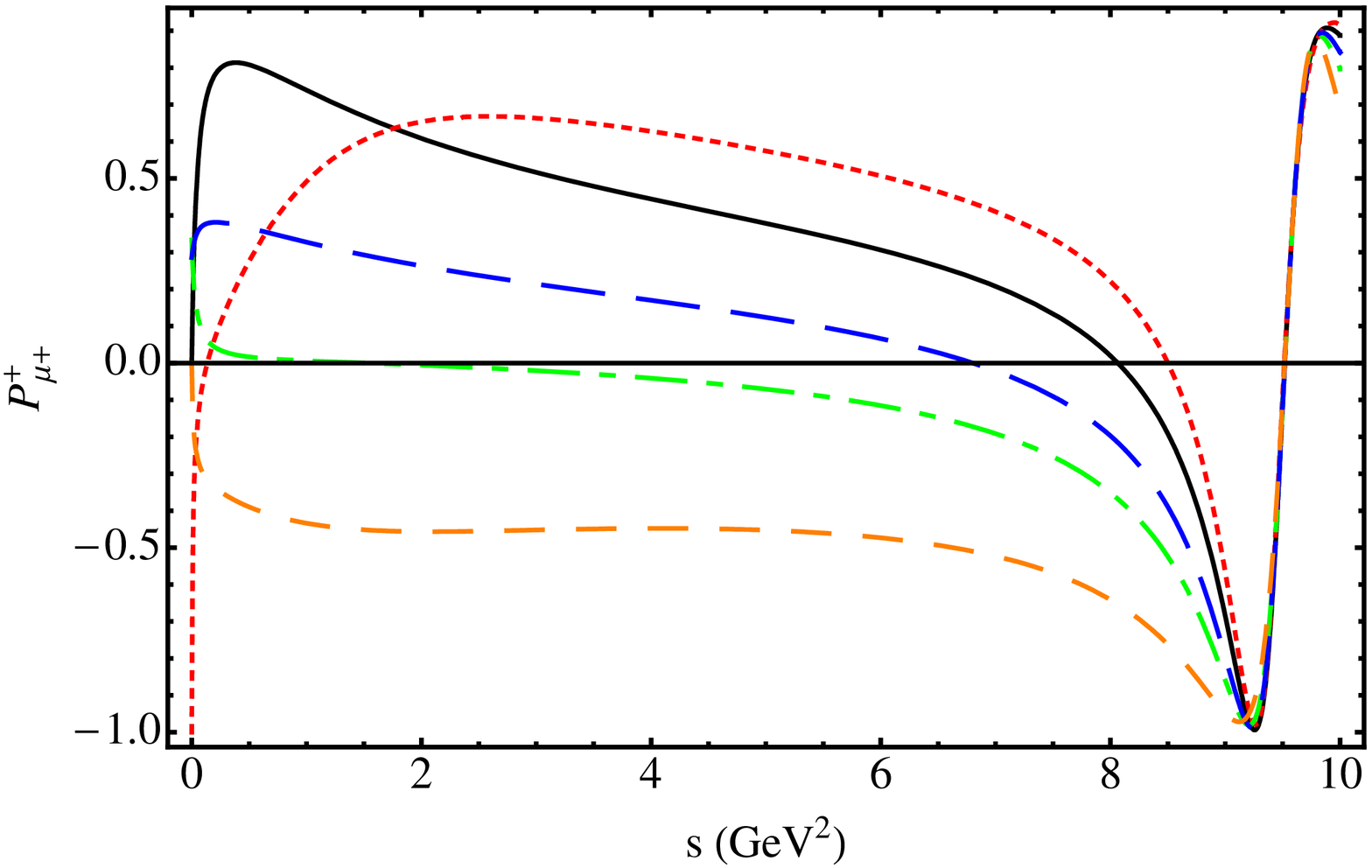,width=.45\textwidth}
\caption{\sl (Colour online) The normalised single muon polarisation asymmetry (${\cal P}_{\mu^-}^+$) for a range of values for $C_{7L}$ and $C_{7R}$.: $C_{7L}/C_{7R} = 0$ (Black) solid line, $C_{7L}/C_{7R} = +1$ (Red) dotted line, $C_{7L}/C_{7R} = -1$ (Green) dot-dashed line, $C_{7L}/C_{7R} = 0.5$ (Blue) dashed line, and $C_{7R}/C_{7L} = 0$ (Orange) small dashed line. The top left panel is for $y = 0$, $z = 0$, the top right panel for $y = +5$, $z = -10$, whilst the bottom left panel is for $y= +5$, $z = -7.5$ and the bottom right panel is for $y = +10$ and $z= -5$.}
\label{fig:3}
\end{center}
\end{figure}

 With this parameterisation we calculate the various lepton polarisation asymmetries and plot in figures \ref{fig:1} to \ref{fig:3} for some typical values of $(x,u,v)$, and give the zeroes of these plots in table \ref{tab:one}\footnote{For ${\cal P}_{\mu^-}^+(y=5,z=7.5) = 0$ we only present the zeroes near $s=2(GeV)^2$}. The results show sensitivity to the values of $x$, $y$ and $z$ quite clearly, where data on the dileptonic decay mode of the ${\bar B} \to {\bar K}_2(1430) \, \ell^+ \, \ell^-$ would be very useful in testing physics BSM. Note that we have restricted our attention to the large recoil region, where we expect the LEET to be more valid. The zeroes of the various lepton polarisation asymmetries, which is a crucial index for comparison with experimental data \cite{Ali:1999mm}, fortunately falls well within this region for most choices of the BSM parameters. We expect that experimental measurements of this dileptonic mode will be available in the near future and the comparison of those with the present theoretical estimates would provide a very useful complement to the corresponding analysis of the well established $K^*(890)$ case. 

\par Furthermore, a paper recently appeared from the Belle collaboration \cite{Wei:2009zv} and has given indications of new physics BSM, which could be of the type suggested by our equation (\ref{eqn:18}) with non-zero phases. Therefore, it would be interesting to also experimentally measure the lepton polarisation asymmetries in ${\bar B} \to {\bar K}_2(1430)$ transitions, such as we are proposing. For the decay of $B \to K^*(890) \, \ell^+ \, \ell^-$, the Belle result had only 230 events. The number of events into the corresponding $K_2(1430)$ may be expected to be about half this number making the statistics even more limited. However, the statistics would drastically improve when data from LHCb becomes available. Estimates made in reference \cite{Rutz} for the LHCb collaboration show that the $B \to K^*(890) \, \ell^+ \, \ell^-$ would have about 8000 events annually, making the analysis of the FB asymmetry more definitive. The corresponding decay channel $ B \to K_2(1430) \, \ell^+ \, \ell^-$ considered here would also have a few thousand events making for meaningful comparisons with the SM and theories beyond possible.

%%%%%%%%%%%%%%%%%%%%%%%%%%%%%%%%%%%%%
%  Acknowledgements

\section*{Acknowledgements}\label{sec:ack}
SRC would like to acknowledge the Department of Science and Technology, Government of India, for support through a research project. 

%%%%%%%%%%%%%%%%%%%%%%%%%%%%%%%%%%%%%
%  Appendices

\appendix

\section{The form factors}\label{appendix:a}

\par We shall take the form factors $U_1$, $U_2$ and $U_3$  from Cheng {\it et al.} \cite{Cheng:2004yj}; the remaining form factors can be related to these using the relations in Charles {\it et al.} \cite{Charles:1998dr} as spelt out earlier:
\bea
U_1 (s) & = & \frac{0.19}{1 - 2.22 (s/m_B^2) + 2.13 (s/m_B^2)^2} \,\,\, , \nonumber \\
U_2 (s) & = & \frac{0.19}{\left( 1 - s/m_B^2 \right) \left( 1 - 1.77 (s/m_B^2) + 4.32 (s/m_B^2)^2 \right)} \,\,\, , \nonumber \\
U_3 (s) & = & \frac{0.16}{1 - 2.19 (s/m_B^2) + 1.80 (s/m_B^2)^2} \,\,\, . \nonumber
\eea

\section{Input parameters and Wilson coefficients}\label{appendix:b}

\par The input parameters used in the generation of the numerical results are as follows:
\begin{center}
$m_B = 5.26$GeV ~~,~~ $m_{K^*} = 1.43$GeV ~~,~~ $m_b = 4.8$GeV ~~,~~ $m_c = 1.4$GeV , \\
$m_s = 0.1$GeV ~~,~~ ${\cal B}(J/\psi(1S) \to \ell^+ \ell^-) = 6 \times 10^{-2}$ , \\ $m_{J/\psi(1S)} = 3.097$GeV ~~,~~ ${\cal B}(\psi(2S) \to \ell^+ \ell^-) = 8.3 \times 10^{-3}$ , \\$m_{\psi(2S)} = 3.097$GeV ~~,~~ $\Gamma_{\psi(2S)} = 0.277 \times 10^{-3}$GeV , \\ $\Gamma_{J/\psi(1S)} = 0.088 \times 10^{-3}$GeV ~~,~~ $V_{tb} V_{ts}^* = 0.0385$ ~~,~~ $\alpha = \frac{1}{129}$ ~~,~~ $G_F = 1.17 \times 10^{-5}$ GeV$^{-2}$.
\end{center}

\noindent The Wilson coefficients used were as in Kim {\it et al.} \cite{Kim:2000dq}, namely:
\bea
C_{7L} & = & - \sqrt{0.081} \cos x \exp\left( i (u+v)\right) \,\,\, , \nonumber \\
C_{7R} & = & - \sqrt{0.081} \sin x \exp\left( i (u-v)\right) \,\,\, , \nonumber \\
C_{10} &=& - 4.546 \,\,\, , \nonumber
\eea
\bea
C_9^{SM} & = & 4.153 + 0.381 g\left(\frac{m_c}{m_b} , \frac{s}{m_B^2} \right) + 0.033 g\left(1, \frac{s}{m_B^2} \right) + 0.032 g\left(0, \frac{s}{m_B^2} \right) - 0.381 \times 2.3 \times \frac{3 \pi}{\alpha} \nonumber \\
&& \hspace{1 cm} \times \left( \frac{\Gamma_{\psi(2S)} {\cal B}(\psi(2S) \to \ell^+ \ell^-) m_{\psi(2S)}}{s - m_{\psi(2S)}^2 + i m_{\psi(2S)} \Gamma_{\psi(2S)}} + \frac{\Gamma_{J/\psi(1S)} {\cal B}(J/\psi(1S) \to \ell^+ \ell^-) m_{J/\psi(1S)}}{s - m_{J/\psi(1S)}^2 + i m_{J/\psi(1S)} \Gamma_{J/\psi(1S)}}\right) , \nonumber
\eea
where the function $g$ is taken from reference \cite{gfunc}:
\bea
g(\hat{m}_i, \hat{s}) & = & - \frac{8}{9} \ln (\hat{m}_i) + \frac{8}{27} + \frac{4}{9} \left( \frac{4 \hat{m}_i^2}{\hat{s}} \right) - \frac{2}{9} \left( 2 + \frac{4 \hat{m}_i^2}{\hat{s}} \right) \sqrt{ \left| 1 - \frac{4 \hat{m}_i^2}{\hat{s}} \right|} \nonumber \\
&& \hspace{1 in} \times \left\{ \begin{array}{lc}
\left| \ln \left(\frac{1 + \sqrt{1 - 4 \hat{m}_i^2/\hat{s}}}{1 - \sqrt{1 - 4 \hat{m}_i^2/\hat{s}}} \right) - i \pi \right| & , 4 \hat{m}_i^2 < \hat{s} \\
2 \arctan \frac{1}{\sqrt{4 \hat{m}_i^2/\hat{s} - 1}} & , 4 \hat{m}_i^2 > \hat{s}
\end{array} \right. \,\,\, . \nonumber
\eea

%%%%%%%%%%%%%%%%%%%%%%%%%%%%%%%%%%%%%
%%%%%%%%%%%%%%%%%%%%%%%%%%%%%%%%%%%%%

%%%%%%%%%%%%%%%%%%%%%%%%%%%%%%%%%%%%%
%%%%%%%%%%%%%%%%%%%%%%%%%%%%%%%%%%%%%

%%%%%%%%%%%%%%%%%%%%%%%%%%%%%%%%%%%%%
%%%%%%%%%%%%%%%%%%%%%%%%%%%%%%%%%%%%%

%%%%%%%%%%%%%%%%%%%%%%%%%%%%%%%%%%%%%
%  Bibliography

\end{document}